\documentclass[journal]{IEEEtran}
\ifCLASSINFOpdf \else \fi
\usepackage{graphicx}
\usepackage[cmex10]{amsmath}
\usepackage{amssymb}
\usepackage{bm}
\usepackage{algorithm}
\usepackage{algpseudocode}

\usepackage{array}
\usepackage{threeparttable}
\usepackage{multirow}
\usepackage{multicol}
\usepackage{diagbox}
\usepackage[keeplastbox]{flushend} 
\ifCLASSOPTIONcompsoc
 \usepackage[caption=false,font=normalsize,labelfont=sf,textfont=sf]{subfig}
\else
 \usepackage[caption=false,font=footnotesize]{subfig}
\fi
\usepackage{url}
\usepackage[noadjust]{cite}

\begin{document}
\title{Modified EP MIMO Detection Algorithm with Deep Learning Parameters Selection}
\author{{Hang Chen, Guoqiang Yao, Jianhao Hu, \emph{Member IEEE}}
\thanks{The authors are with the Department of National Key Laboratory of Science and Technology on Communication, University of Electronic Science and Technology of China, Chengdu, Sichuan, 611731 China (e-mail:chenhang2018,guoqiangyao@std.uestc.edu.cn;jhhu@uestc.edu.cn).}}
\markboth{}
{Shell \MakeLowercase{\textit{et al.}}: Bare Demo of IEEEtran.cls for Journals}
\maketitle
\begin{abstract}
Expectation Propagation (EP)-based Multiple-Input Multiple-Output (MIMO) detector is regarded as a state-of-the-art MIMO detector because of its exceptional performance. However, we find that the EP MIMO detector cannot guarantee to achieve the optimal performance due to the empirical parameter selection, including initial variance and damping factors. According to the influence of the moment matching and parameter selection for the performance of the EP MIMO detector, we propose a modified EP MIMO detector (MEPD). In order to obtain the optimal initial variance and damping factors, we adopt a deep learning scheme, in which we unfold the iterative processing of MEPD to establish MEPNet for parameters training. The simulation results show that MEPD with off-line trained parameters outperforms the original one in various MIMO scenarios. Besides, the proposed MEPD with deep learning parameters selection is more robust than EPD in practical scenarios.
\end{abstract}


\IEEEpeerreviewmaketitle

\section{Introduction}

\IEEEPARstart{M}{ultiple-Input} Multiple-Output (MIMO) technology has been adopted in many wireless communication systems \cite{A1OverviewMassiveMIMO}. Massive MIMO is regarded as one of the key technologies in the 5th generation(5G) and future mobile communication system \cite{A2MIMOin5G} because MIMO technology can achieve higher spectral efficiency by transmitting multiple data streams concurrently within the same frequency band. But the detection complexity and energy consumption could be too huge to be acceptable in the Massive MIMO systems. Therefore, the high-performance low-complexity receiver design is of great significance.

\par MIMO technologies have been studied for more than fifty years, and the Massive MIMO systems have become a hot topic of wireless communications \cite{A3FiftyYears}. Due to co-channel interference (CCI), The Massive MIMO systems face great technical challenges in symbols detection. In Massive MIMO symbols detection, the complexity of the conventional optimal detector, the maximum likelihood detector (MLD), is not acceptable. Sphere Decoding (SD) \cite{A4aSD,A4FSD} with sub-optimal performance was proposed to reduce the complexity of MLD by restricting the search space to a high-dimensional sphere, but its complexity is still high and therefore not practical in the Massive MIMO systems. Linear MIMO symbols detectors include matched filter (MF), zero-forcing (ZF), linear minimum mean square error (LMMSE) \cite{A5MassiveMIMOSurvey}. They are adopted in the practical MIMO systems due to their low complexities. However, there are huge performance gaps compared with that of MLD. Several aided methods can be adopted in ZF and MMSE detector to improve their performance with acceptable increased complexity \cite{A6SQRD,A7LRA,A8SIC,A9PIC}.

\par The factor graph-based message passing algorithm for MIMO detection can achieve excellent performance. i.e. Belief Propagation (BP), Expectation Propagation (EP). EP-based MIMO detector (EPD) was proposed in \cite{B0EPD} with excellent performance and moderate complexity. EPD approximates and updates the posterior probability of the transmitted symbol vector iteratively on the factor graph using Gaussian approximation and Message Passing (MP). According to our studies, we find that the updating scheme of moment matching in EPD limits its performance and the parameters which are empirically determined are not optimal. To the authors’ best knowledge, there is no optimal parameters selection criterion and it’s hard to find such criterion mathematically, so we adopt the deep learning scheme to optimize the parameters.

\par In the past decades, Deep Learning (DL) has been applied to various fields of scientific research and engineering applications due to its powerful learning ability \cite{B1Deeplearning}. And more recently, it has been applied to the physical layer of communication systems for channel estimation \cite{B2DeepChannelEstimation}, \cite{B3surveydeepchannelestimation}, automatic modulation classification (AMC) \cite{B4AMC}, signal detection \cite{B5DataSequencesDetection}, \cite{B8MMNet}, channel decoding \cite{B9DeepLinearCodes}, \cite{C0TurboNet}, and etc. These successful applications illustrate that deep learning can be applied to the physical layer of communication systems and achieve excellent performance \cite{C1ondeepdecoding,C1opticalMIMO,C2end2end,C31DeepLearningAided,C4DNNSDconstraint,C5DNNSDprediction,C3Deepunfolding,B6DeepMIMOdetection,B7LearningtoDetect}. The authors of papers \cite{C6OAMPNet,C7OAMPNet2} have developed a model-driven DL network for symbols detection in the MIMO systems, named OAMPNet, which is obtained by unfolding the orthogonal approximate message passing (OAMP) \cite{C8OAMP} algorithm, an improved approximate message passing (AMP) \cite{C8aAMP1,C8bAMP2} algorithm, and adding four trainable parameters. Its essence is an optimization method based on training. Simulation results showed that the OAMPNet detector significantly improves the performance compared to the OAMP detector, and exhibits robustness to various mismatches. It also illustrates the advantages of optimization methods based on deep learning scheme.

\par In this paper, we propose a modified EP-based MIMO detector (MEPD), in which a modified updating scheme of moment matching is adopted. MEPD algorithm has similar computational complexity with EPD, but we focus on the excellent performance of MEPD. In order to determine the optimal parameters, including initial variance and damping factors, we unfold MEPD and name it MEPNet, then the optimal parameters can be determined through training. The key differences between this work with DL-SEP \cite{C8DL-SEP} are that: First, we propose a modified updating scheme of moment matching which can make full use of the self-correction ability of the EP algorithm, the simplified form is more suitable for training. Second, initial variance which can accelerate convergence if carefully selected is optimized through off-line training. The main contributions are listed as follows:

\begin{itemize}
  \item A modified updating scheme of moment matching which can make full use of the self-correction ability of EP algorithm is proposed. The practical significance of a negative "variance" is detailed. To improve the detection performance of EPD, the modified updating scheme of moment matching insists on the updating process when a negative "variance" appears and utilizes the self-correction property of the EP algorithm adequately.
  \item We propose a modified EP-based MIMO detection (MEPD) algorithm that surpasses EPD in SER performance and robustness in practical scenarios. The proposed MEPD algorithm takes in consideration that the optimal value of key parameters vary with noise power and iteration.
  \item A model-based data-driven neural network named MEPNet is proposed for parameters selection. Since there is no rigorous mathematic criterion for parameters selection, we adopt a deep learning scheme to finish it. Specifically, these parameters to be determined are set as trainable variables in MEPNet established by unfolding MEPD algorithm, such that the optimal values of these parameters can be obtained by training MEPNet.
\end{itemize}

\par The rest of this paper is organized as follows: System model and the EP-based detection algorithm are briefly described in Section II. In Section III, the modified EP MIMO detector is proposed based on performance studies. In Section IV, the structure of MEPNet established for parameters optimization and analyses of training results are presented. Then, simulation results are detailed in Section V. Finally, Section VI concludes the paper.

\par \emph{Notations}: The lowercase and uppercase in boldface denote column vector and matrix, respectively. $(\mathbf{A})^{T}$ , $(\mathbf{A})^{-1}$,$\mathbf{A}(i,j)$ denote the transpose,  inversion, the entry at $i^{th}$ row $j^{th}$column of matrix $\mathbf{A}$, respectively. A superscript with brackets denotes the iteration or layer index. A subscript denotes the item index of a vector. $\mathbb{R}(\cdot)$ and $\mathbb{I}(\cdot)$ are the real part and imaginary part of a complex number, vector, or matrix. $\mathbb{E}_{p(a)}[a]$ and $\mathbb{V}_{p(a)}[a]$ denote the expectation and variance of random variable $a$, the probability density function (PDF) of $a$ is $p(a)$. $\mathcal{C} \mathcal{N}(\mathbf{x}:\mathbf{a},\mathbf{B})$ and $\mathcal{C} \mathcal{N}(x:a,b)$ denote complex joint Gaussian and complex Gaussian random variables, where the three parameters are random variable(s), mean(s) and (co)variance(matrix), respectively. Similarly, $\mathcal{N}(\mathbf{x}:\mathbf{a},\mathbf{B})$ and $\mathcal{N}(x:a,b)$ denote real joint Gaussian and Gaussian random variables. The operation $\operatorname{diag}(\cdot)$ acts on a vector will get a diagonal matrix.

\section{Preliminaries}

\subsection{System Model}
\par The MIMO system can be simply formulated as:
\begin{equation}
\bar{\mathbf{y}} = \bar{\mathbf{H}}\bar{\mathbf{x}} + \bar{\mathbf{n}}\label{eq1}
\end{equation}
where $\bar{\mathbf{y}} \in \mathbb{C}^{N_{r} \times 1}, \bar{\mathbf{H}} \in \mathbb{C}^{N_{r} \times N_{t}}, \bar{\mathbf{x}} \in \mathbb{C}^{N_{t} \times 1}, \bar{\mathbf{n}} \in \mathbb{C}^{N_{r} \times 1}$ are the received symbol vector, channel matrix, transmitted symbol vector and system noise vector in the complex domain, respectively. $N_{t}$ and $N_{r}$ are the numbers of transmitting and receiving antennas.

\par We assume that $\bar{\mathbf{H}}$ is unitary invariant, $\bar{\mathbf{n}}$ is additive white Gaussian noise (AWGN) whose elements follows $\mathcal{C}\mathcal{N}(0,\sigma_{\bar{n}}^{2})$. $\bar{\mathbf{x}}$ is randomly and uniformly selected from the constellation set $\bar{\Theta}$. The average energy for transmitted symbols is:

\begin{equation}
E_{\bar{x}}=\frac{1}{|\bar{\Theta}|} \sum_{k=1}^{|\bar{\Theta}|}\left\|\bar{\theta}_{k}\right\|^{2}\label{eq2}
\end{equation}
where $\bar{\Theta}=\left\{\bar{\theta}_{1}, \bar{\theta}_{2}, \ldots, \bar{\theta}_{|\bar{\Theta}|}\right\}$, $|\bar{\Theta}|$ denotes the constellation size and $\left\|\bar{\theta}_{k}\right\|$ is modulus of constellation point $\bar{\theta}_{k}$. The signal to noise ratio (SNR) is:
\begin{equation}
SNR = 10\log_{10}\frac{N_{t}E_{\bar{x}}}{\sigma_{\bar{n}}^{2}}\label{eq3}
\end{equation}

\par When it comes to Bayesian inference, the calculation of posterior probability distributions is usually performed as real-valued. Thus, the complex-valued system model is transformed into real-valued by $\mathbf{y}=\left[\mathbb{R}(\bar{\mathbf{y}})^{T} \quad \mathbb{I}(\bar{\mathbf{y}})^{T}\right]^{T}, \mathbf{x}=\left[\mathbb{R}(\bar{\mathbf{x}})^{T} \quad \mathbb{I}(\bar{\mathbf{x}})^{T}\right]^{T}, \mathbf{n}=$
$\left[\mathbb{R}(\bar{\mathbf{n}})^{T} \quad \mathbb{I}(\bar{\mathbf{n}})^{T}\right]^{T}$, and
\begin{equation}
\mathbf{H}=\left[\begin{array}{cc}
\mathbb{R}(\bar{\mathbf{H}}) & -\mathbb{I}(\bar{\mathbf{H}}) \\
\mathbb{I}(\bar{\mathbf{H}}) & \mathbb{R}(\bar{\mathbf{H}})
\end{array}\right]\label{eq4}
\end{equation}

\par Then the real-valued model is expressed as:

\begin{equation}
\mathbf{y} = \mathbf{H}\mathbf{x} + \mathbf{n}\label{eq5}
\end{equation}

\par In the real-valued model Eq. \eqref{eq5}, the average signal energy $E_{x}=0.5E_{\bar{x}}$, noise power $\sigma_{n^{2}}=0.5\sigma_{\bar{n}^{2}}$. Let $N=2N_{t}$, $M=2N_{r}$, there are $\mathbf{y} \in \mathbb{R}^{M \times 1}$, $\mathbf{H} \in \mathbb{R}^{M \times N}$, $\mathbf{x} \in \mathbb{R}^{N \times 1}$, $\mathbf{n} \in \mathbb{R}^{M \times 1}$.

\subsection{EPD Algorithm}
\par EPD approximates and updates the posterior probability of the transmitted symbol vector iteratively on the factor graph using Gaussian approximation and message passing. The posterior probability distribution (PoPD) for the transmitted symbol vector in the MIMO systems can be expressed as:
\begin{equation}
p(\mathbf{x} | \mathbf{y,H}) = \frac{p(\mathbf{x},\mathbf{y}|\mathbf{H})}{p(\mathbf{y})}=\frac{p(\mathbf{y}|\mathbf{x,H})p(\mathbf{x})}{p(\mathbf{y})}\propto p(\mathbf{y}|\mathbf{x,H})p(\mathbf{x})\label{eq6}
\end{equation}
where $p(\mathbf{y}|\mathbf{x,H})$ follows $\mathcal{N}(\mathbf{y}:\mathbf{H}\mathbf{x},\sigma_{n}^{2}\mathbf{I}_{\mathbf{M}})$, and $p(\mathbf{x})=\prod_{i=1}^{N}p(x_{i})$. The EP algorithm makes an assumption that the marginal prior probability distributions (MPrPDs) of transmitted symbols are $\hat{p}(x_{i})=\operatorname{exp}(-1/2\lambda_{i}x_{i}^{2}+\gamma_{i}x_{i}),i=1,2,\dots,N$, so the approximated PoPD can be expressed as:
\begin{equation}
\begin{array}{l}
\hat{p}(\mathbf{x} | \mathbf{y,H}) \propto \mathcal{N}\left(\mathbf{y}: \mathbf{H} \mathbf{x}, \sigma_{n}^{2} \mathbf{I}_{M}\right) \cdot \prod_{i=1}^{N} \hat{p}\left(x_{i}\right) \\
\propto \exp \left(-\frac{(\mathbf{y}-\mathbf{H} \mathbf{x})^{T} (\mathbf{y}-\mathbf{H} \mathbf{x})}{2 \sigma_{n}^{2}}\right) \prod_{i=1}^{N} \exp \left(-\frac{1}{2} \lambda_{i} x_{i}^{2}+\gamma_{i} x_{i}\right) \\
=\exp \left(-\frac{(\mathbf{y}-\mathbf{H} \mathbf{x})^{T} (\mathbf{y}-\mathbf{H} \mathbf{x})}{2 \sigma_{n}^{2}}+\sum_{i=1}^{N}\left(-\frac{1}{2} \lambda_{i} x_{i}^{2}+\gamma_{i} x_{i}\right)\right)
\end{array}\label{eq7}
\end{equation}
since the distribution is a product of Gaussian distributions, it also follows Gaussian distribution:
\begin{equation}
\hat{p}(\mathbf{x} | \mathbf{y,H}) \sim \mathcal{N}(\mathbf{x}: \mathbf{u}, \mathbf{C}) \propto \exp \left(-\frac{1}{2}(\mathbf{x}-\mathbf{u})^{T} \mathbf{C}^{-1}(\mathbf{x}-\mathbf{u})\right)\label{eq8}
\end{equation}
where $\mathbf{C}$ and $\mathbf{u}$ are the covariance matrix and mean vector of approximated PoPD, respectively. Comparing Eq. \eqref{eq7} with Eq. \eqref{eq8}, there are:
\begin{equation}
\mathbf{C}=\left(\sigma_{n}^{-2} \mathbf{H}^{T} \mathbf{H}+\operatorname{diag}(\boldsymbol{\lambda})\right)^{-1}\label{eq9}
\end{equation}
\begin{equation}
\mathbf{u}=\mathbf{C} \cdot\left(\sigma_{n}^{-2} \mathbf{H}^{T} \mathbf{y}+\bm{\gamma}\right)\label{eq10}
\end{equation}
where $\bm{\lambda}=[\lambda_{1},\lambda_{2},\dots,\lambda_{N}]^{T}$, $\bm{\gamma}=[\gamma_{1},\gamma_{2},\dots,\gamma_{N}]^{T}$. Then the rest process of EPD in one iteration is to update $\lambda_{i}$ and $\gamma_{i}$ of approximated MPrPD by non-linear estimation (NLE). The iterative  processing of EPD terminates until the EP algorithm converges or maximum iterations is performed. The EPD algorithm is summarized as Algorithm \ref{EPD}.

\begin{algorithm}
	\caption{EPD Algorithm}
	\label{EPD}
	\begin{algorithmic}[1]
		\Require
		$\mathbf{H}$, $\mathbf{y}$, $\sigma_n^2$, $E_x$, $N$, $Iter$
		\Ensure
		$\hat{\mathbf{x}}$
		\State Initialization: $\lambda_i^{(1)} = 1/E_x$, $\gamma_i^{(1)}=0$, $\epsilon = 5\times 10^{-7}$, $\beta=0.2$, $t=1$;
		\While{$t \leq Iter$}
		\State $\bm{\lambda}^{(t)} = [\lambda_1^{(t)},\lambda_2^{(t)},...,\lambda_{N}^{(t)}]^T$;
		\State $\bm{\gamma}^{(t)} = [\gamma_1^{(t)},\gamma_2^{(t)},...,\gamma_{N}^{(t)}]^T$;
		\State $\mathbf{C}^{(t)} = \left(\sigma_n^{-2}\mathbf{H}^T\mathbf{H} + \mathit{diag} (\bm{\lambda}^{(t)})\right)^{-1}$; \label{line5}
		\State $\mathbf{u}^{(t)} = \mathbf{C}^{(t)} \cdot \left(\sigma_n^{-2} \mathbf{H}^T \mathbf{y} + \bm{\gamma}^{(t)} \right)$;
		\For{$i=1 \to N$}   $\backslash\backslash$ \textit{Parallel Execution}
		\State $e_i=\mathbf{u}^{(t)}(i)$, $s_i=\mathbf{C}^{(t)}(i,i)$;
		\State 	$\varepsilon_i^2 = \frac{s_i}{1-s_i \cdot \lambda_i^{(t)}}$; \label{line9}
		\State $m_i=\varepsilon_i^2 \left(\frac{e_i}{s_i} - \gamma_i^{(t)} \right)$;\label{line10}
		\State $u_i^{\prime} = \mathbb{E}_{p(\theta)}\left[\theta\right]$; \label{line11}
		\State $\sigma_i^{\prime 2} = \max\left( \epsilon, \mathbb{V}_{p(\theta)}\left[\theta\right] \right)$; \label{line12}
		\If{$\sigma_i^{\prime 2} > \varepsilon_i^{2}$} \label{line13}
		\State  $\lambda_{i}^{(t+1)} = \lambda_{i}^{(t)}$; \label{line14}
		\State  $\gamma_{i}^{(t+1)} = \gamma_{i}^{(t)}$;   \label{line15}
		\Else
		\State $\lambda_{i}^{(t+1)}=(1-\beta)\lambda_i^{(t)} + \beta\left( \frac{1}{\sigma_i^{\prime 2}}-\frac{1}{\varepsilon_i^{2}}\right)$; \label{line17}
		\State $\gamma_{i}^{(t+1)}= (1-\beta)\gamma_i^{(t)} + \beta\left( \frac{u_i^{\prime}}{\sigma_i^{\prime 2}} - \frac{m_i}{\varepsilon_i^{2}} \right)$;\label{line18}
		\EndIf \label{line19}
		\EndFor
		\State $t = t + 1$;
		\EndWhile
		\State $\hat{\mathbf{x}} = \mathbf{u}^{(t)} $;
	\end{algorithmic}
\end{algorithm}
Where $\theta \in \Theta=\left\{\theta_{1}, \theta_{2}, \ldots, \theta_{|\Theta|}\right\}$ is the real-valued constellation point, and the PDF of $\theta$ is:
\begin{equation}
p(\theta)=\frac{\mathcal{N}\left(\theta: m_{i},\varepsilon_{i}^{2}\right) \delta\left(\theta-\theta_{j}\right)}{\sum_{j=1}^{|\Theta|} \mathcal{N}\left(\theta: m_{i},\varepsilon_{i}^{2}\right) \delta\left(\theta-\theta_{j}\right)}\label{eq11}
\end{equation}
$\delta(z)$ is the Dirac function which is equal to 1 when $z=0$ otherwise $0$.

\subsection{Limitation and Motivation}
\par EP algorithm is proved to be Bayes-optimal in the large system limit \cite{C9EP-SE}, where both input and output dimensions tend to infinity while the compression rate is kept constant. But when the EP algorithm adopted into MIMO symbols detection, the practical number of antennas could not be infinite and the ratio($\eta=N/M$) between the number of transmitting and receiving antennas does not necessarily satisfy a threshold. Namely, only when $M\rightarrow\infty$ and $\eta$ satisfy the compression rate threshold, EPD can achieve Bayes-optimal performance. We believe that the performance of EPD can be improved in the practical MIMO systems.

\section{MEPD Algorithm}
\par In this section, we propose a modified EP MIMO detector (MEPD). First, by illustrating the practical significance of a negative variance,  we adopt a modified updating scheme of moment matching in EPD to achieve better performance. Then we analyze two factors that can impact the performance of EPD with the modified updating scheme of moment matching. Finally, the MEPD algorithm is presented according to our analysis.

\subsection{Updating Scheme of Moment Matching}
\begin{figure}[!t]
      \centering
      \includegraphics[width= 3.5 in]{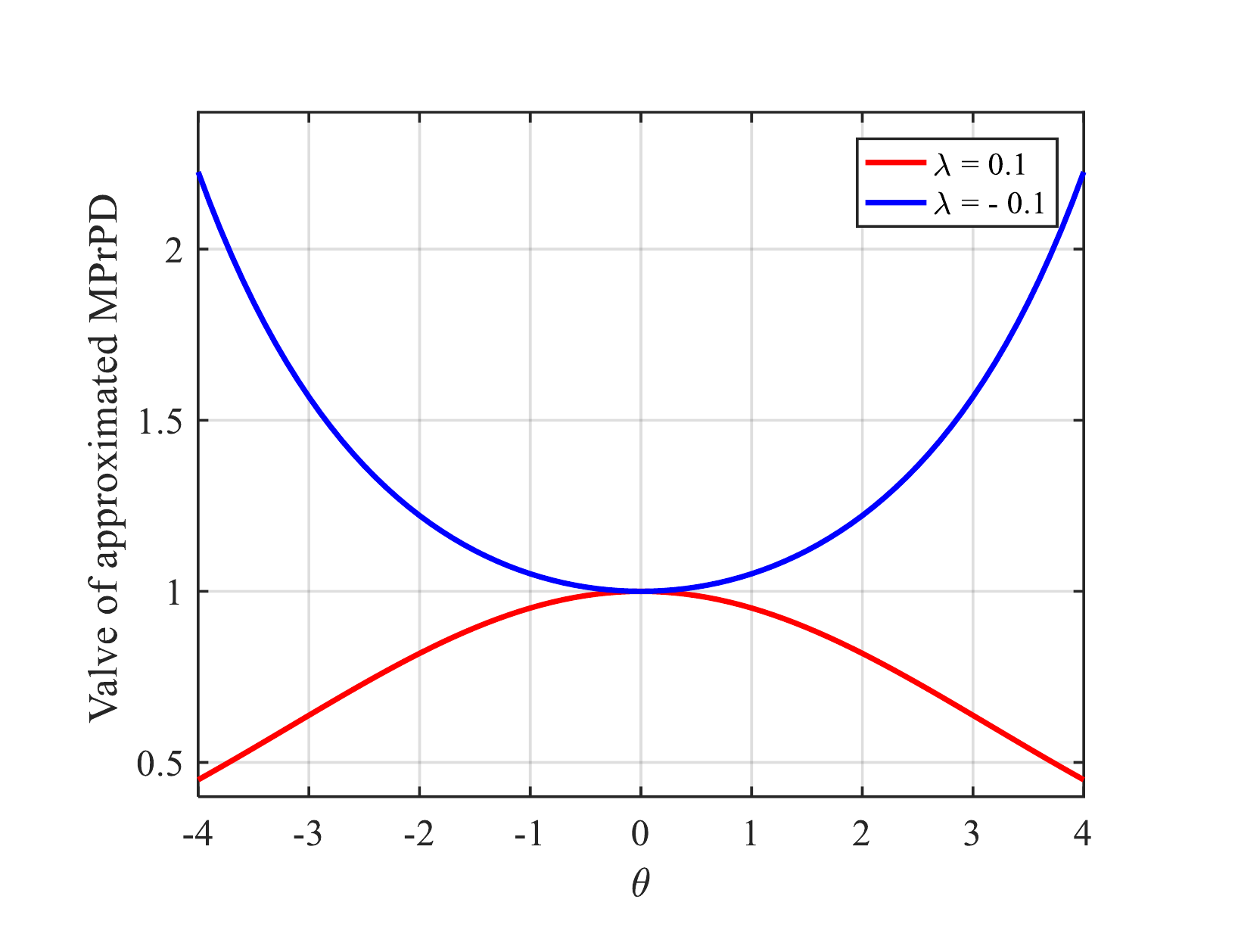}
      \caption{\small{The sampling values of the approximated marginal prior distributions with positive or negative parameter $(\lambda = ±0.1)$ and zero-mean $(\gamma = 0)$}}
      \label{fig1}
\end{figure}

The moment matching can update the means and variances of the approximated MPrPDs, and it is the key step for the updating process of the EP algorithm. \figurename{\ref{fig1}} shows the sampling values of the approximated MPrPDs with positive or negative variance ($\lambda=\pm0.1$) and zero-mean ($\gamma=0$).

\par From \figurename{\ref{fig1}}, a negative variance indicates that the probability of the sampling value at the current mean is the lowest. In other words, the positive variance indicates the present mean is the optimal approximation for the transmitted symbol while the negative variance represents the approximation is the worst. When an approximated mean is far away from the constellations, the updated variance could be negative. Therefore, a negative $\lambda$ propagates the following information to the next iteration: the present approximation is inaccurate. It is also trying to correct the inaccuracy by updating a negative value.

\par Thus, the operation of ignoring the present updating will lose the information which the corresponding moment matching propagates, and it will restrain the self-correction ability of the EP algorithm, and consequently the performance of EPD. Therefore, the proposed updating scheme insists on the updating process when a negative $\lambda$ appears and utilizes the self-correction property of the EP algorithm adequately. It is worth mentioning that this scheme has been directly given in \cite{C9aEPNSA,C9bEPNSA}, but the reason is not explained.

\par Lines \ref{line13} to \ref{line19} in Algorithm \ref{EPD} can be simplified to Eq. \eqref{eq12} and Eq. \eqref{eq13} because in the proposed updating scheme of moment matching we do not ignore a negative variance.
\begin{equation}
\lambda_{i}^{(t+1)}=(1-\beta)\lambda_i^{(t)} + \beta\left( \frac{1}{\sigma_i^{\prime 2}}-\frac{1}{\varepsilon_i^{2}}\right)\label{eq12}
\end{equation}
\begin{equation}
\gamma_{i}^{(t+1)}= (1-\beta)\gamma_i^{(t)} + \beta\left( \frac{u_i^{\prime}}{\sigma_i^{\prime 2}} - \frac{m_i}{\varepsilon_i^{2}} \right)\label{eq13}
\end{equation}
Then by combining lines \ref{line9} and \ref{line10} in Algorithm \ref{EPD} and Eq. \eqref{eq12}, Eq. \eqref{eq13}, we get:
\begin{equation}
\lambda_{i}^{(t+1)}= \lambda_{i}^{(t)}+\beta\left(\frac{1}{\sigma_{i}^{\prime 2}}-\frac{1}{s_i}\right)\label{eq14}
\end{equation}
\begin{equation}
\gamma_{i}^{(t+1)}=\gamma_{i}^{(t)}+\beta\left(\frac{u_{i}^{\prime}}{\sigma_{i}^{\prime 2}}-\frac{e_i}{s_i}\right)\label{eq15}
\end{equation}

\par
\begin{figure}[!t]
  \centering
  \includegraphics[width=3.5 in]{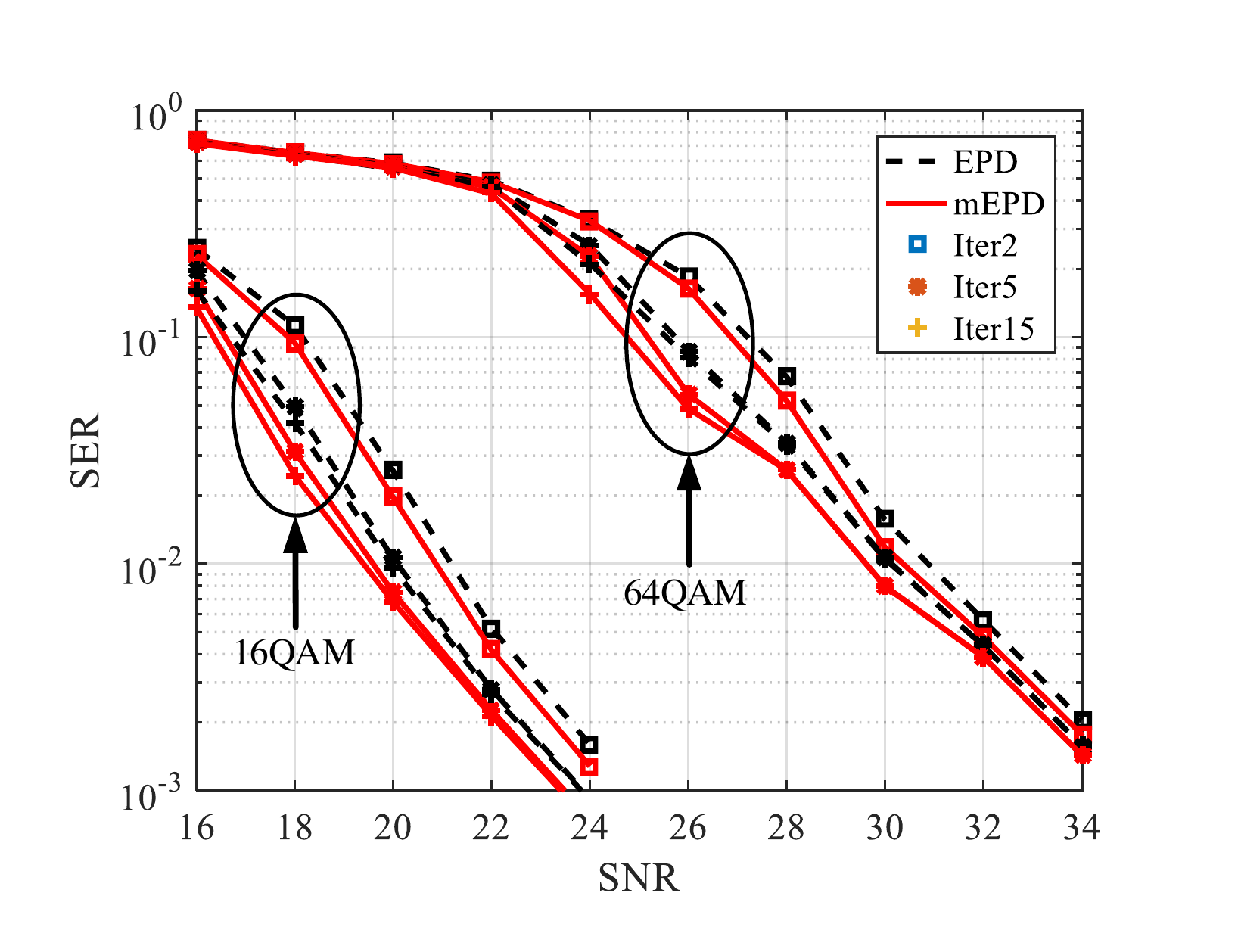}
  \caption{\small{SERs performance of the mEPD and EPD versus SNRs in $N_t=N_r=32$ MIMO with 16QAM and 64QAM modulation.}}
  \label{fig2}
\end{figure}

 For the sake of distinction, EPD with the modified updating scheme of moment matching is named mEPD. Simulation experiments in \figurename{\ref{fig2}} illustrate the correctness of our analysis. Although the performance gain of mEPD is not great.

\subsection{Initial Variance and Damping Factor}
\par In this work, we find two factors listed as follows can impact the performance of mEPD (Indeed, EPD has the same problems. Due to space limitations, we only analyze mEPD and then give an optimization scheme):
\begin{itemize}
    \item[1.] mEPD gives LMMSE solution in first iteration by setting initial parameters $\lambda_i^{(1)}=1/E_x$, $\gamma_i^{(1)}=0$;
    \item[2.] During the iteration of mEPD, the damping factor is set to a constant of 0.2, which means that mEPD believes that the newly calculated value has a lower and fixed confidence.
\end{itemize}

\par We call the abovementioned two factors initial variance and damping factor, respectively. And in the rest of this section we present how the two factors impact the performance of mEPD.

\subsubsection{Initial Variance}
\par mEPD adopts initial parameter $\lambda_i^{(1)}=1/E_x=0.2$ when 16QAM modulation, which reaches the optimal SER performance in the first iteration. Because mEPD with $\lambda_i^{(1)}=\lambda = 0.2$ is equal to the LMMSE equalization. For simplicity, we assume the value of $\lambda_i^{(1)}$ is $\lambda$.

\begin{figure}[!t]
      \centering
      \includegraphics[width= 3.5 in]{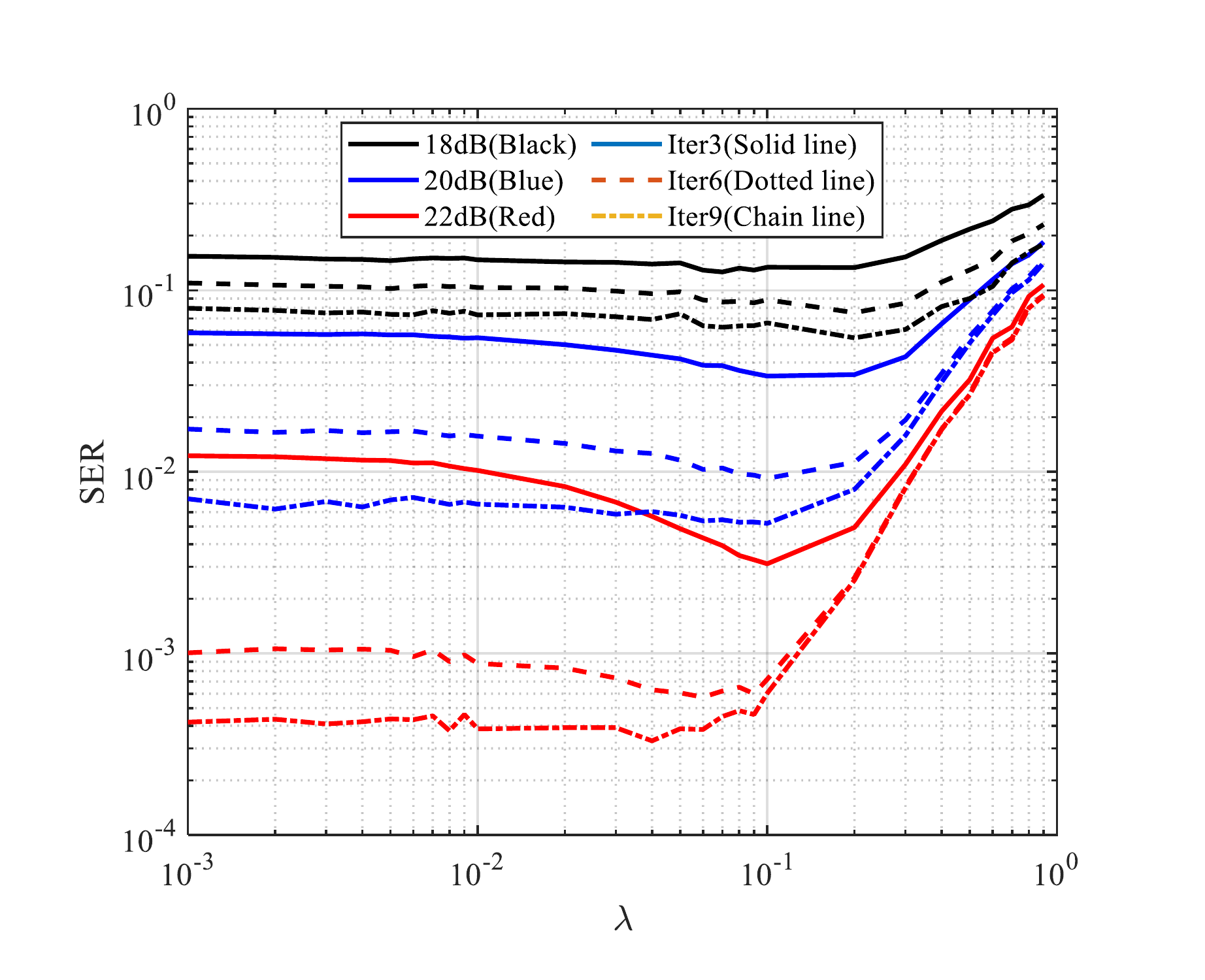}
      \caption{\small{SERs performance of mEPD with $\beta=0.2$ versus $\lambda$ in $N_t = N_r = 16$ MIMO with 16QAM modulation}}
      \label{fig3}
\end{figure}

\par However, with the increasing of iterations, the optimal SER performance will no longer be obtained when $\lambda=0.2$. \figurename{\ref{fig3}} shows that when SNR $= 18$ dB, $0.2$ is the optimal value of $\lambda$, while the optimal value of $\lambda$ is 0.1 when SNR $= 20$ dB. But when SNR comes to $22$ dB, the optimal value of $\lambda$ varies with iterations. That is, a larger $\lambda$ results in faster convergence, but a smaller $\lambda$ results in better performance. The same results can be found in other scenarios, due to space limitations, it will not be shown here. Thus, $\lambda=1/E_x$ is not the optimal initialization scheme for mEPD.
\par To summarize, under the condition that the number of antennas and the modulation order are unchanged, the optimal initial variances are different when the number of iterations and the SNRs are different.

\subsubsection{Damping Factor}
To improve the robustness, The EP algorithm was suggested to adopt a low-pass filter to smooth the moment update. Hence, mEPD employs a damping scheme to update the posterior moments as illustrated in Algorithm \ref{EPD} lines \ref{line17} and \ref{line18}. It’s worth noting that the EP algorithm is a message passing algorithm (MPA), a fixed $\beta$ means that mEPD reckons newly computed posterior moments have a fixed and lower confidence in all iterations and SNRs. It is unreasonable.

\begin{figure}[!t]
  \centering
  \includegraphics[width=3.5 in]{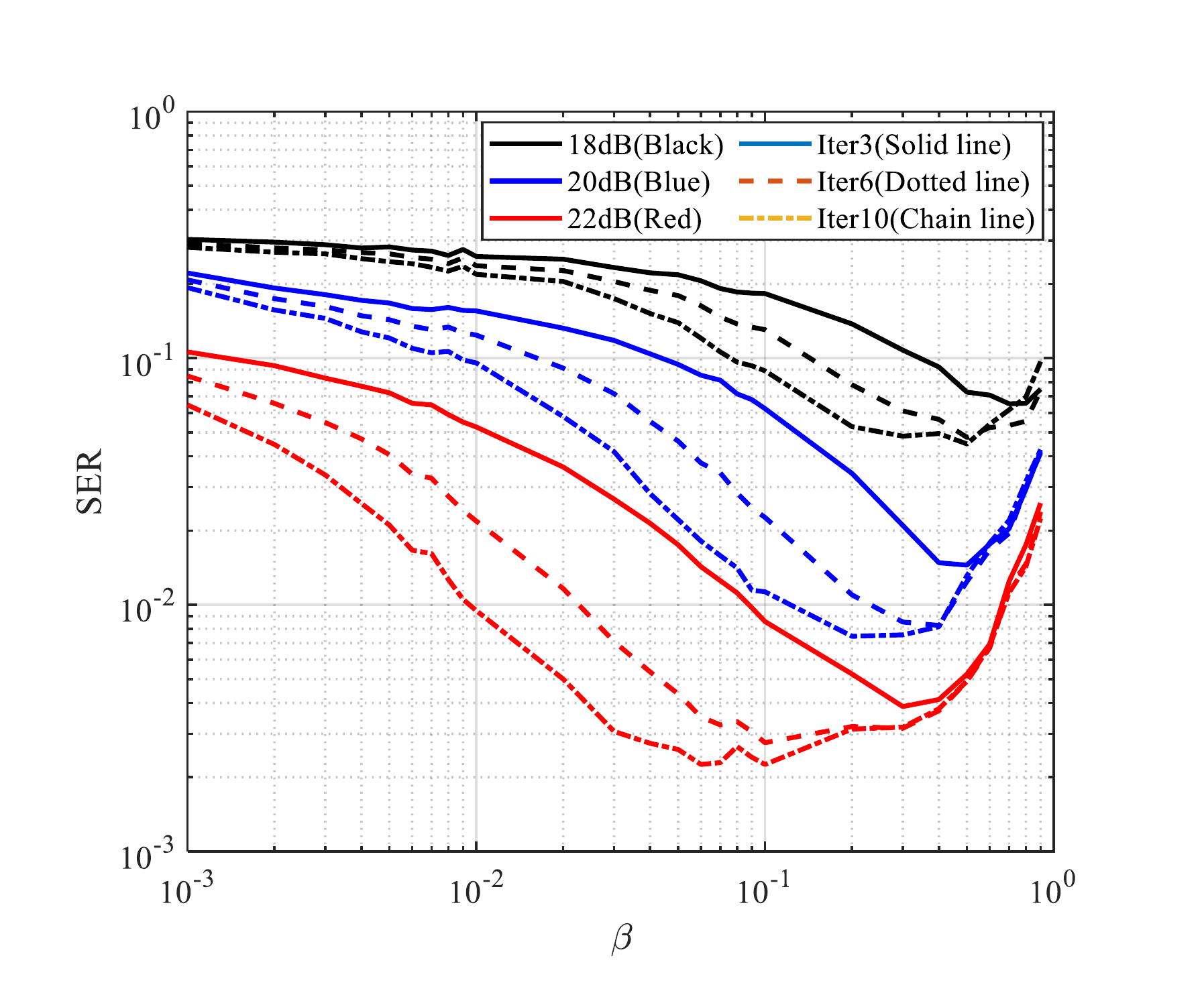}
  \caption{\small{SERs performance of mEPD when $\lambda=1/E_x$ versus $\beta$ in $N_t = N_r = 16$ MIMO with 16QAM modulation}}
  \label{fig4}
\end{figure}

\par As shown in \figurename{\ref{fig4}}, the optimal $\beta' s$ have different values under different SNRs and number of iterations. We can find that the trend of the optimal $\beta$ is similar with that of $\lambda$, a larger $\beta$ results in faster convergence, a smaller $\beta$ results in better performance but needs more iterations, consequently leads to higher complexity.

\subsection{MEPD Algorithm}

\par In the previous subsections, we demonstrated that ignoring negative variance limits the performance of EPD. To extract the EP algorithm, we proposed mEPD. Then we illustrated that initial variance and damping factor in mEPD are not optimal for all cases. More specifically, a larger $\beta$ or $\lambda$ results in faster convergence, a smaller $\beta$ or $\lambda$ will result in better performance but needs more iterations, and their optimal values vary with SNRs and iteration times.
\par A question comes that can mEPD achieve better performance as well as fast convergence speed by choosing optimal parameters. For this purpose, we propose MEPD for MIMO detection summarized in Algorithm 2. Obviously, MEPD algorithm has the similar computation complexity with EPD.

\begin{algorithm}
	\caption{MEPD Algorithm}
	\label{OMEPD}
	\begin{algorithmic}[1]
		\Require
		$\mathbf{H}$, $\mathbf{y}$, $\sigma_n^2$, $\bm{\alpha}$, $\bm{\beta}$, $\lambda$, $N$, $Iter$
		\Ensure
		$\hat{\mathbf{x}}$
		\State Initialization: $\lambda_i^{(1)} = \lambda$, $\gamma_i^{(1)}=0$, $\epsilon = 5\times 10^{-7}$, $t=1$;
		\While{$t \leq Iter$}
		\State $\bm{\lambda}^{(t)} = [\lambda_1^{(t)},\lambda_2^{(t)},...,\lambda_{N}^{(t)}]^T$;
		\State $\bm{\gamma}^{(t)} = [\gamma_1^{(t)},\gamma_2^{(t)},...,\gamma_{N}^{(t)}]^T$;
		\State $\mathbf{C}^{(t)} = \left(\sigma_n^{-2}\mathbf{H}^T\mathbf{H} + \mathit{diag} (\bm{\lambda}^{(t)})\right)^{-1}$; \label{l5}
		\State $\mathbf{u}^{(t)} = \mathbf{C}^{(t)} \cdot \left(\sigma_n^{-2} \mathbf{H}^T \mathbf{y} + \bm{\gamma}^{(t)} \right)$;
		\For{$i=1 \to N$}   $\backslash\backslash$ \textit{Parallel Execution}
		\State $e_i=\mathbf{u}^{(t)}(i)$, $s_i=\mathbf{C}^{(t)}(i,i)$;
		\State 	$\varepsilon_i^2 = \frac{s_i}{1-s_i \cdot \lambda_i^{(t)}}$; \label{l9}
		\State $m_i=\varepsilon_i^2 \left(\frac{e_i}{s_i} - \gamma_i^{(t)} \right)$;\label{l10}
		\State $u_i^{\prime} = \mathbb{E}_{p^{\prime}(\theta)}\left[\theta\right]$; \label{l11}
		\State $\sigma_i^{\prime 2} = \max\left( \epsilon, \mathbb{V}_{p^{\prime}(\theta)}\left[\theta\right] \right)$; \label{l12}
		\State $\lambda_{i}^{(t+1)}= \lambda_{i}^{(t)}+\beta_t\left(\frac{1}{\sigma_{i}^{\prime 2}}-\frac{1}{s_i}\right)$; \label{l17}
		\State $\gamma_{i}^{(t+1)}=\gamma_{i}^{(t)}+\beta_t\left(\frac{u_{i}^{\prime}}{\sigma_{i}^{\prime 2}}-\frac{e_i}{s_i}\right)$;\label{l18}
		\EndFor
		\State $t = t + 1$;
		\EndWhile
		\State $\hat{\mathbf{x}} = \mathbf{u}^{(t)} $;
	\end{algorithmic}
\end{algorithm}

In Algorithm \ref{OMEPD}, $\bm{\alpha}=[\alpha_1,\alpha_2,\dots,\alpha_{Iter}]^T$, $\bm{\beta}=[\beta_1,\beta_2,\dots,\beta_{Iter}]^T$, and $\lambda$, are the optimal parameters matched with estimated noise power $\sigma_n^2$. $\epsilon$ is a small positive constant by following the convention that substituting $\sigma_i^{\prime 2} = \mathbb{V}_{p^{\prime}(\theta)}\left[\theta \right]$ by $\sigma_i^{\prime 2} = \max\left( \epsilon, \mathbb{V}_{p^{\prime}(\theta)}\left[\theta \right] \right)$ to avoid stability problem. The PDF of $\theta$ is modified to be:
\begin{equation}
p^{\prime}(\theta)=\frac{\mathcal{N}\left(\theta: m_{i},\alpha_t\varepsilon_{i}^{2}\right) \delta\left(\theta-\theta_{j}\right)}{\sum_{j=1}^{|\Theta|} \mathcal{N}\left(\theta: m_{i},\alpha_t\varepsilon_{i}^{2}\right) \delta\left(\theta-\theta_{j}\right)}\label{eq16}
\end{equation}
where $\alpha_t$ is added to adjust the variances of cavity distributions for better performance. The main differences between MEPD and EPD are: 1) the updating scheme of moment matching is modified to take full advantage of the self-correction ability of EP algorithm; 2) the variances of cavity distributions are scaled; 3) initial variance is not $1/E_x$ and damping factors are different in each iteration, their optimal values are pre-determined. When $\alpha_t=1$, $\beta_t=0.2$, and $\lambda=1/E_x$, MEPD is reduced to mEPD.
\par But the key problem has not been solved because the optimal parameters have not been determined. Obviously, number of optimal parameters is $2*Iter+1$. So what we are solving now is a multivariate optimization problem. In the next section, we will introduce an optimization scheme based on training.

\section{MEPNet}
\par In this section, we introduce the MEPNet which can learn the optimal parameters for MEPD, then the training scheme and practical implementation scheme are presented. Finally, the analyses about optimized parameters are given at the end of this section.

\subsection{The Structure of MEPNet}

\begin{figure*}[!t]
      \centering
      \includegraphics[width= 5.5 in]{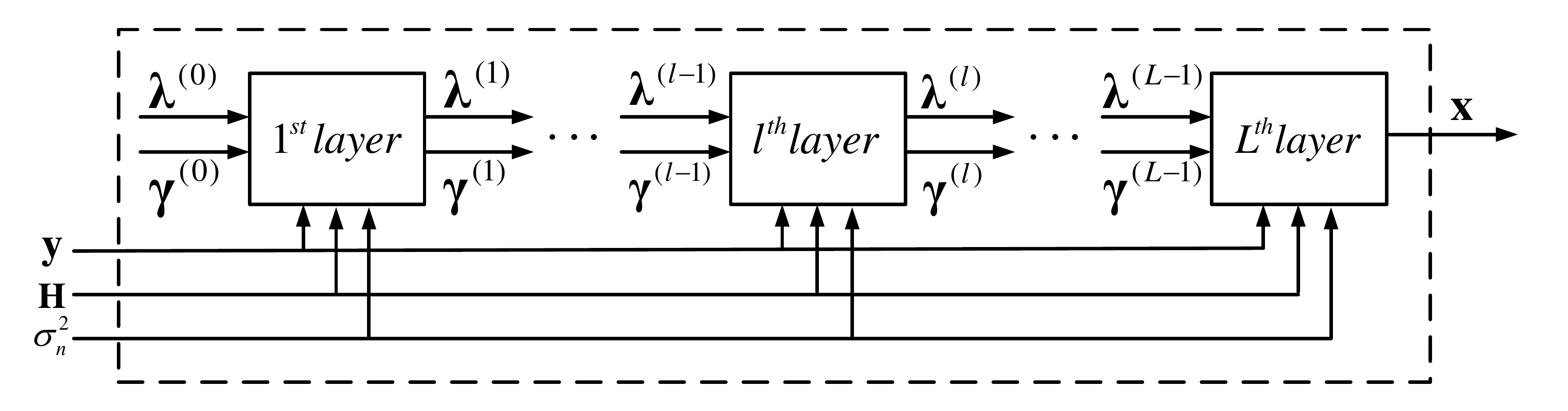}
      \caption{\small{The block diagram of MEPNet. The network consists of L cascade layers, and each layer has the same structure}}
      \label{fig5}
\end{figure*}
\par The structure of MEPNet is illustrated in \figurename{\ref{fig5}}, which is obtained by unfolding MEPD and selecting learnable variables ($\bm{\alpha},\bm{\beta},\lambda$). In the rest of this section, we assume the layer index of the input and output of $l^{th}$ layer is $(l-1)$ and $(l)$, respectively. The input of MEPNet is received signal vector, $\mathbf{y}$, channel matrix, $\mathbf{H}$, and estimated noise power, $\hat{\sigma_n^{2}}$, while the final output is the estimation of transmitted symbol vector $\hat{\mathbf{x}}$. The network consists of L layers, and each layer has the same structure with 2 learnable variables(3 in first layer). In $l^{th}$ layer, MEPNet performs the posterior probability approximation as lines 3 to 15 in Algorithm 2.

\par It seems that EPD can also be trained by a DL scheme. However, there are two drawbacks of it: 1) MEPD with empirical parameters, $\alpha_t=1$, $\beta_t=0.2$, and $\lambda=1/E_x$, outperforms EPD as shown in \figurename{\ref{fig2}}; 2) By comparing lines \ref{line13} to \ref{line19} in Algorithm \ref{EPD} and lines 13 and 14 in Algorithm 2, we can find that the updating scheme of moment matching in MEPD is more suitable for training and easier to converge, because the comparison operation in EPD may result in gradient problems in training.

\par Obviously, the number of learnable variables is $2L+1$ ($L$ is the number of layers) of MEPNet. By adjusting these learnable variables in training stage, we can greatly improve the detection performance of MEPD compared to EPD.

\subsection{Training Scheme and Practical Implementation Scheme}
\begin{figure}[!t]
      \centering
      \includegraphics[width= 3.5 in]{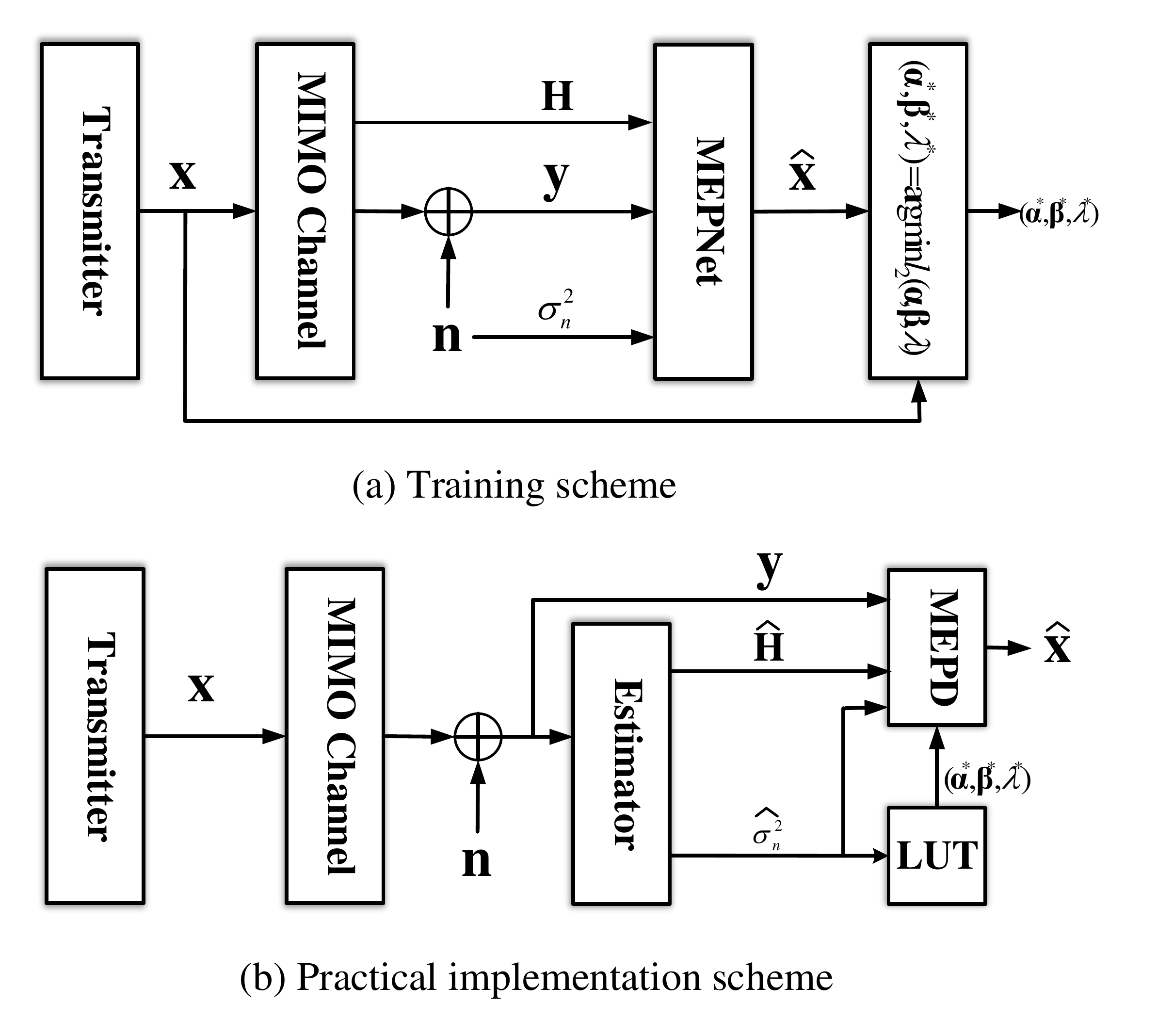}
      \caption{\small{Block diagram of (a) training scheme of MEPNet and (b) practical implementation scheme of MEPD}}
      \label{fig6}
\end{figure}

\par The training scheme of this work is as shown in \figurename{\ref{fig6}}(a), the transmitted symbol vector $\mathbf{x}$ is randomly selected in the constellation set $\Theta$ in the training stage, the i.i.d. Rayleigh MIMO channel $\mathbf{H}$ is randomly generated while its complex-valued form $\bar{\mathbf{H}}(j,i)\sim \mathcal{C}\mathcal{N}(0,1)$ and the noise vector is randomly generated whose element is $\bar{n}_j \sim \mathcal{C}\mathcal{N}(0,\sigma_{\bar{n}}^2)$. We use perfect channel state information in training stage, then we train MEPNet until all learnable variables converge.

\par Based on the analysis in Section III, the optimal parameters are different under different antenna scales and modulation orders at different SNRs, therefore, we train MEPNet at different system configuration and SNRs. $(\bm{\alpha}^{\star},\bm{\beta}^{\star},\lambda^{\star})$ represent optimized parameters and $l_2(\bm{\alpha},\bm{\beta},\lambda)$ is the cost function used in training defined as:
\begin{equation}
l_{2}(\boldsymbol{\alpha}, \boldsymbol{\beta}, \lambda)=\frac{1}{|\pi|} \sum_{n=1}^{|\pi|}\left\|\mathbf{x}_{n}-\widehat{\mathbf{x}}_{n}\right\|^{2}
\label{eq24}
\end{equation}
where $\pi$ is mini-batch of training data pairs with batch size of $|\pi|$.
\par We also propose a practical implementation scheme as shown in \figurename{\ref{fig6}}(b). After training process finished, all optimized parameters at every SNR can be stored, then MEPD can automatically load the corresponding optimal parameters according to the MIMO system configuration and estimated noise power($\hat{\sigma_n^2}$) while assuming that channel matrix is perfectly estimated, namely $\hat{\mathbf{H}}=\mathbf{H}$. We will present the robust of MEPD for the mismatch of $\hat{\sigma_n^2}$ and channel correlation in practical scenarios.

\begin{figure}[!t]
      \centering
      \includegraphics[width= 3.5 in]{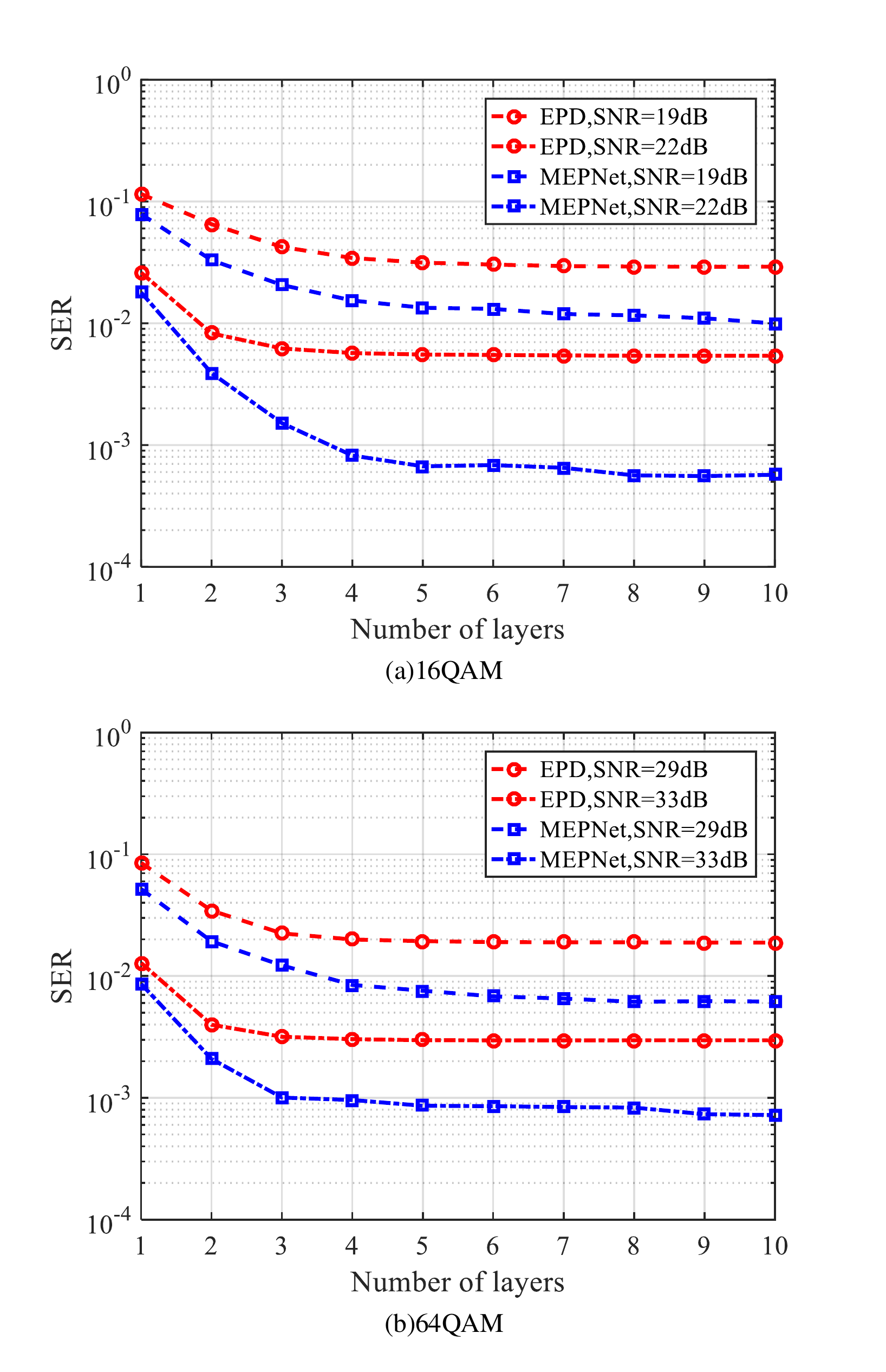}
      \caption{\small{SERs performance of the MEPD and EPD versus the number of layers under (a)16QAM and (b)64QAM modulation.}}
      \label{fig7}
\end{figure}

\par \figurename{\ref{fig7}} illustrates the SER performance of MEPD and EPD versus the number of layers(iterations) in $32\times32$ MIMO with 16QAM and 64QAM modulation. We set the number of layers(iterations) $L=5$ in the following simulations. Because from \figurename{\ref{fig7}} we can find that EPD converges within 5 iterations and MEPD can obtain nearly the best performance at 5 layers, therefore $L=5$ is reasonable by considering that the detection complexity of MEPD better not be higher than that of EPD. In addition, \figurename{\ref{fig7}}(a) shows that MEPD can obtain larger performance gain when SNR $=22$ dB compared with that of SNR $=19$ dB under 16QAM modulation, the same result can be seen in \figurename{\ref{fig7}}(b) under 64QAM modulation. That means MEPD can obtain larger performance gain in high SNR regime, and simulation results in section V can also support this point.

\subsection{Analyses of the Optimized Parameters}
\begin{figure}
  \centering
  \includegraphics[width=3.5in]{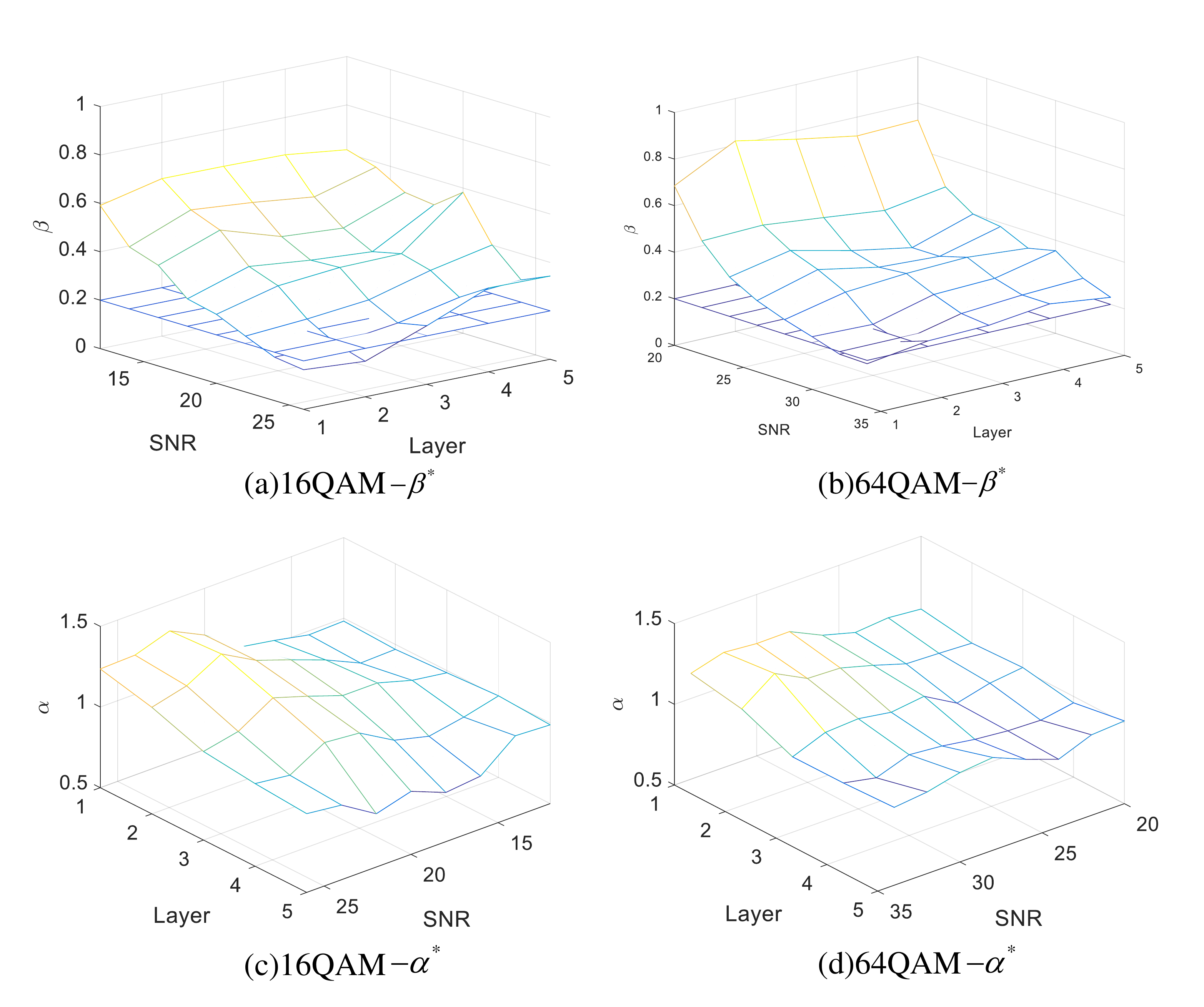}
  \caption{\small{The optimized $\alpha^{\star}$ and $\beta^{\star}$ of MEPD under $32\times32$ i.i.d. Rayleigh MIMO channels.}}
  \label{fig8}
\end{figure}

\renewcommand\arraystretch{2}
\begin{table*}[]
\tiny
\centering
\caption{$\lambda^{\star}$of MEPD under 32$\times$32 i.i.d. Rayleigh MIMO channels}
\label{lambda}
\resizebox{\textwidth}{!}{%
\begin{tabular}{cccccccccc}
\hline
\multirow{2}{*}{16QAM} & SNR/dB            & 12     & 14     & 16     & 18     & 20     & 22     & 24     & 26     \\
                       & $\lambda^{\star}$ & 0.1153 & 0.1137 & 0.1067 & 0.0775 & 0.0651 & 0.0571 & 0.0490 & 0.0369 \\ \hline
\multirow{2}{*}{64QAM} & SNR/dB            & 20     & 22     & 24     & 26     & 28     & 30     & 32     & 34     \\
                       & $\lambda^{\star}$ & 0.0550 & 0.0283 & 0.0243 & 0.0227 & 0.0183 & 0.0136 & 0.0113 & 0.0096 \\ \hline
\end{tabular}%
}
\end{table*}
\par \figurename{\ref{fig8}} and TABLE I illustrate the optimized $\bm{\alpha^{\star}}$, $\bm{\beta^{\star}}$, and $\lambda^{\star}$ of MEPD under $32\times32$ i.i.d. Rayleigh MIMO channels with 16QAM and 64QAM modulation, respectively. In this subsection, we analyze the significance of the optimized parameters.

\subsubsection{$\alpha^{\star}$}
\figurename{\ref{fig8}}(c) and \figurename{\ref{fig8}}(d) illustrate the trend of $\alpha^{\star}$ with the change of layer and SNR. We can find that in the first two layers of high SNR regime $\alpha^{\star}>1$, which means the variance of cavity distribution is scaled larger. An intuitive explanation to this trend is: when SNR is high enough, the results of the first two layers are of high confidence, in order to accelerate convergence, MEPD amplifies the variances of the cavity distributions.

\subsubsection{$\beta^{\star}$}
From \figurename{\ref{fig8}}(a) we can see that the overall trend of $\beta^{\star}$ is becoming larger with the increasing of layer, which means as the iteration progresses, MEPD believes that the confidence of the new prior mean and variance is increasing. Therefore, MEPD can reduce error propagation to a certain extent and in subsequent obtain performance gain. On the SNR axis, $\beta^{\star}$ decreases as SNR becomes larger, because in high SNR regime, the result of the first calculation already has relatively high accuracy. In 64QAM modulation, this trend is more obvious as shown in \figurename{\ref{fig8}}(b). But EPD set $\beta$ the same value in all iterations and it is determined based on simulations under a few scenarios. In addition, almost all $\beta^{\star}$ is larger than 0.2 which is the value of $\beta$ in EPD.

\subsubsection{$\lambda^{\star}$}
The values of $\lambda^{\star}$s of MEPD under $32\times32$ i.i.d. Rayleigh MIMO channels with 16QAM and 64QAM modulation are listed in TABLE I. Obviously, $\lambda^{\star}$ becomes smaller as SNR becomes larger. An intuitive explanation is that when SNR becomes larger, noise power $\sigma_n^2$ is becomes smaller and smaller, the detector is easier to converge to the correct constellation point and a smaller $\lambda$ which means a larger initial variance can accelerate the convergence.

In summary, the training results show that MEPD has a faster convergence speed, which means that MEPD has more excellent performance with less number of iterations. Besides, the differences between neighboring parameters are extremely small, which means MEPD has strong robustness to SNR mismatch. Because if the estimation of system noise power has different degrees of deviation, the parameters matched with $\hat{\sigma_n^2}$ will be loaded into MEPD as shown in \figurename{\ref{fig6}}(b).

\section{Simulation Results}
\par In the previous section, MEPD algorithm is proposed, and its optimal parameters can be determined by training MEPNet which is obtained by unfolding MEPD algorithm. In this section, we provide simulation results to show the performance of MEPD algorithm for MIMO symbols detection.

\subsection{Implementation Details}

\par In the training and validation stage, MEPNet is implemented on the Google Tensorflow platform on a PC with Intel® Core(TM) i5-9600K CPU and NVIDIA GeForce GTX 1050Ti GPU. The training data which is randomly generated consists of a number of data pairs $(\mathbf{x},\mathbf{H},\mathbf{y},\mathbf{n})$, we train MEPNet with 5 layers using Adam optimizer under different antenna scales and modulation orders at different SNRs with 25 epochs. In every epoch, the i.i.d MIMO channel is used, 10,000 data pairs are generated for training with the learning rate is 0.0001 and decaying rate 0.99. We validate the performance every 5 epochs. In the validation stage, we test MEPD until the number of error symbols exceeds 2,000 or the total number of data pairs reaches 100,000.

\par In the correlated MIMO channels used in performance simulation, we use channel matrix condition number to measure channel correlation. It is equivalent to the exponential correlation model \cite{D2Correlated} mathematically because a large correlation coefficient results in a large matrix condition number. The method to generate correlated MIMO channels is presented in Algorithm \ref{GMCMC}. Where the operation $\operatorname{svd}(\cdot)$ denotes the singular value decomposition (SVD) algorithm. $k$ is a constant pre-computed to ensure the condition number of output matrix $\bar{\mathbf{H}}_c$ is $K$, $\bar{\mathbf{H}}$ is an i.i.d Rayleigh MIMO channel whose columns or rows are pre-normalized. Thus, we can provide the performance of MEPD in correlated MIMO channels.

\begin{algorithm}
	\caption{Generation method of correlated MIMO channel}
	\label{GMCMC}
	\begin{algorithmic}[1]
		\Require
		$k$, $\bar{\mathbf{H}}$, $N_t$, $N_r$
		\Ensure
		$\bar{\mathbf{H}}_c$
		\State $[\mathbf{U},\mathbf{S},\mathbf{V}]=\operatorname{svd}(\bar{\mathbf{H}})$;
		\For{$n=1 \to N_t$}
		\State $\mathbf{S}(n,n)=k^{-2(n-1)/N_t}$;
		\EndFor
		\State $\mathbf{S}=N_{t} N_{r} \mathbf{S} / \sum_{n=1}^{N_{t}} \mathbf{S}(n, n)$;
		\State $\bar{\mathbf{H}}_c = \mathbf{U}\mathbf{S}\mathbf{V}$;
	\end{algorithmic}
\end{algorithm}

\subsection{I.I.D MIMO Channel Performance}
\begin{figure}[!t]
      \centering
      \includegraphics[width= 3.5 in]{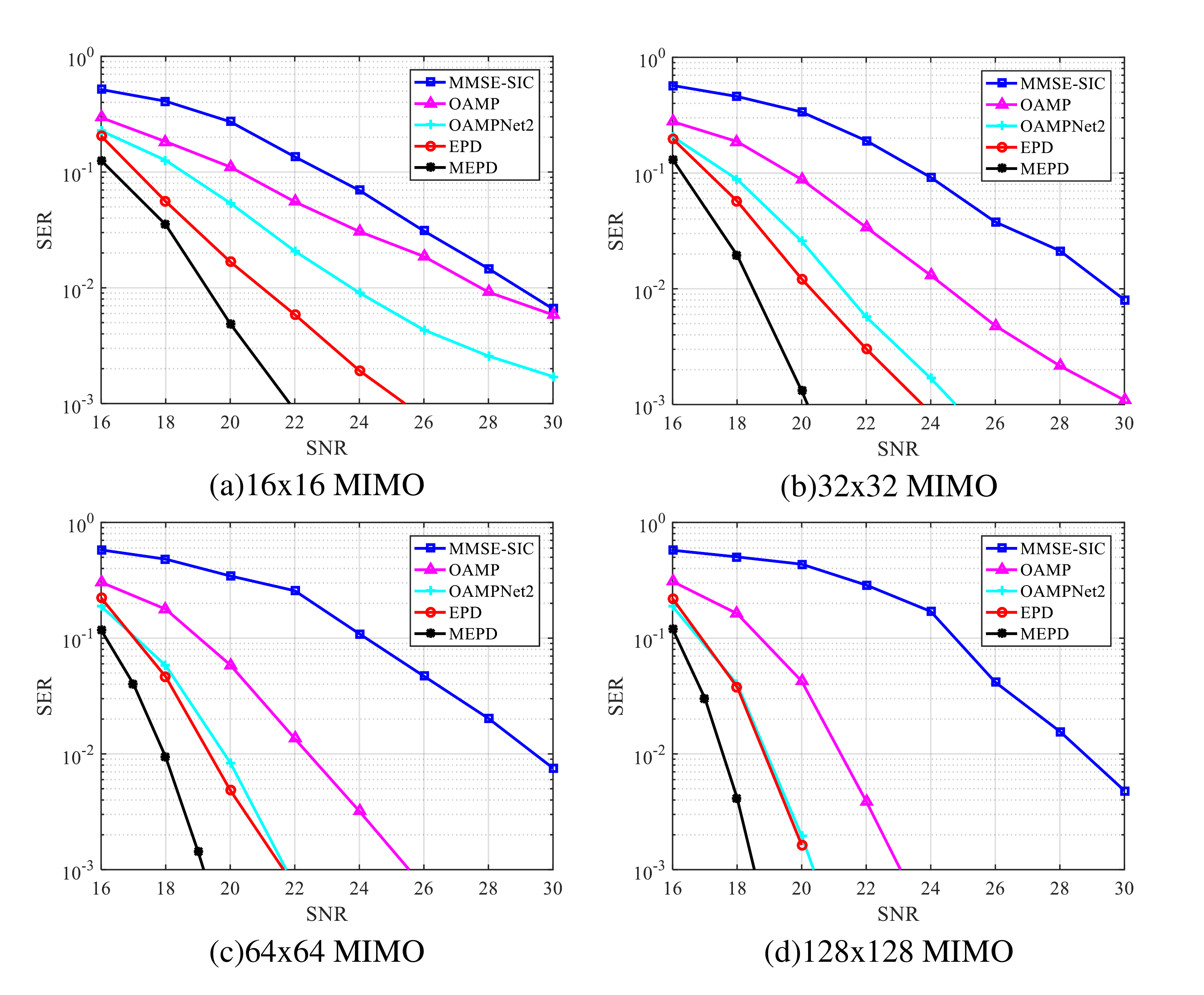}
      \caption{\small{SERs performance of the MEPD and other detectors versus SNRs under i.i.d. Rayleigh MIMO channels with 16QAM modulation.}}
      \label{fig9}
\end{figure}
\begin{figure}[!t]
      \centering
      \includegraphics[width= 3.5 in]{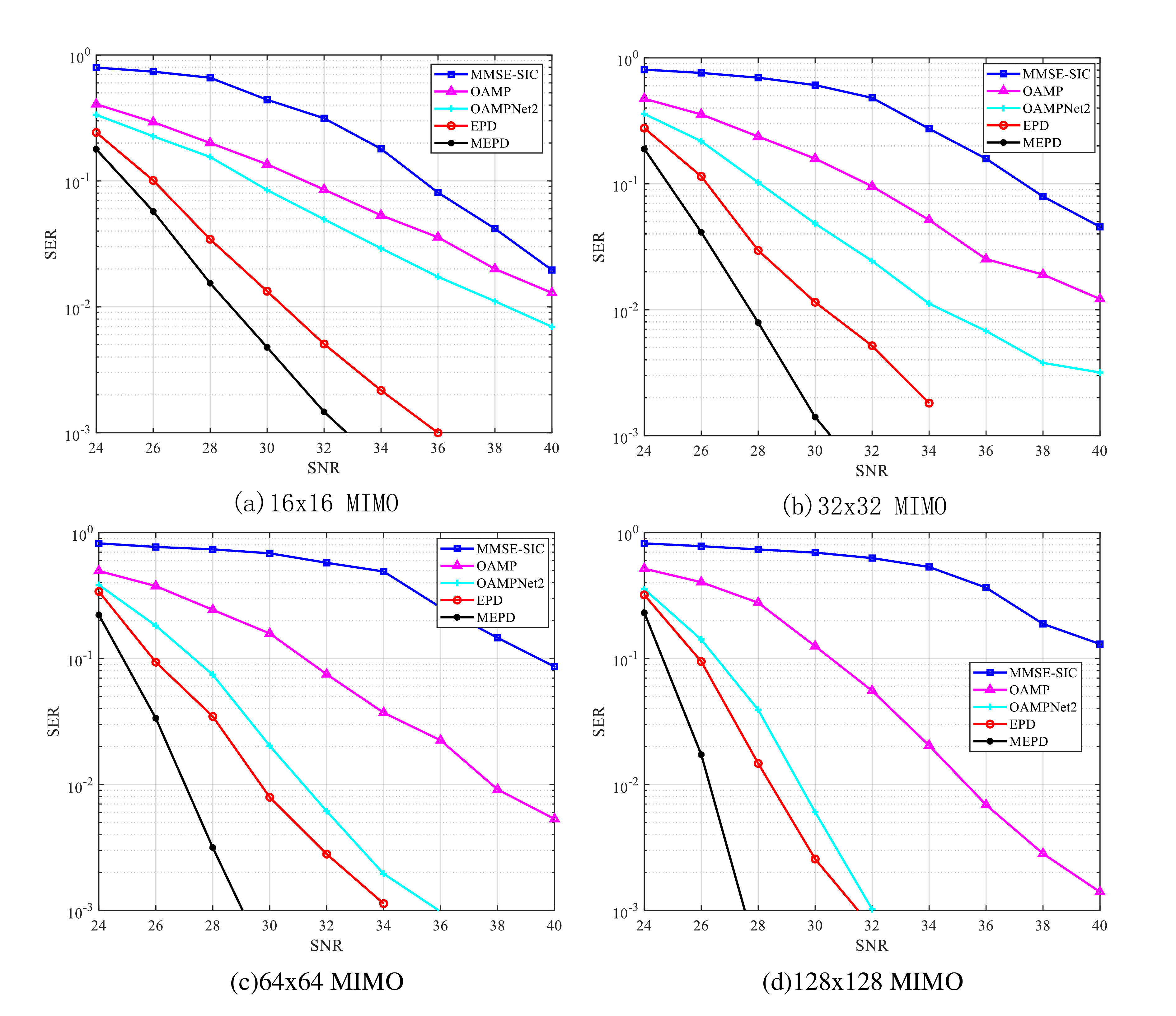}
      \caption{\small{SERs performance of the MEPD and other detectors versus SNRs under i.i.d. Rayleigh MIMO channels with 64QAM modulation.}}
      \label{fig10}
\end{figure}
\begin{figure}[!t]
      \centering
      \includegraphics[width= 3.5 in]{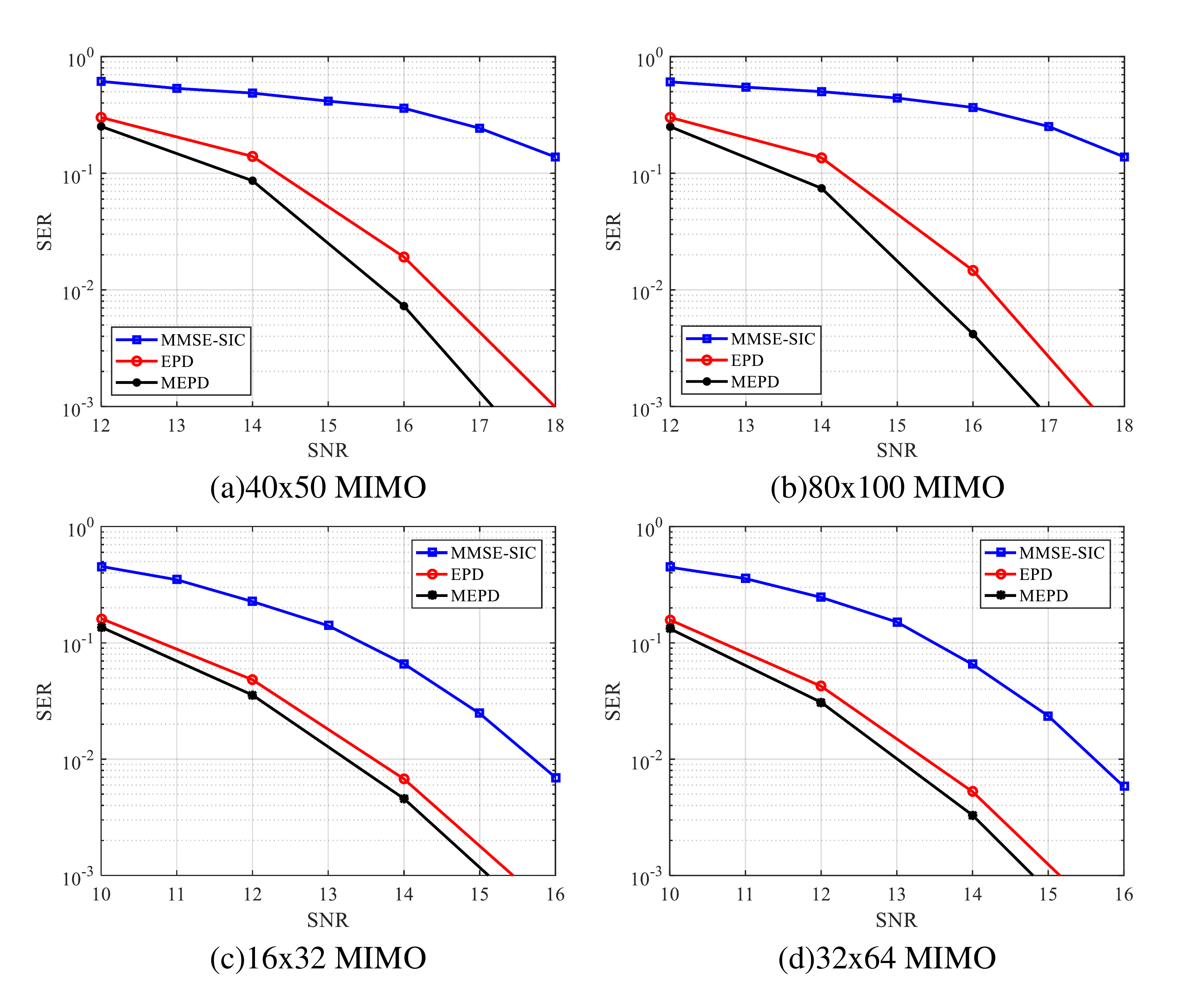}
      \caption{\small{SERs performance of the MEPD and EPD and MMSE-SIC detector versus SNRs under i.i.d. Rayleigh MIMO channels with 16QAM modulation.}}
      \label{fig11}
\end{figure}

\par \figurename{\ref{fig9}} and \figurename{\ref{fig10}} compare the SER performance of MEPD, EPD \cite{B0EPD}, MMSE-SIC \cite{A8SIC}, OAMP \cite{C6OAMPNet}, and OAMPNet2 \cite{C7OAMPNet2} detectors under i.i.d. Rayleigh MIMO channels with 16QAM and 64QAM modulation, respectively. Except for the MMSE-SIC detector, the number of layers or iterations of other detectors are 5. We investigate these detectors in $16\times16$, $32\times32$, $64\times64$, and $128\times128$ MIMO systems. The OAMPNet2 detector is obtained by unfolding the OAMP detector and adding four trainable parameters. In the training process of the OAMPNet2 detector, the training set of each epoch contains 15,000 (20,000 in high SNR regime) data pairs which are randomly generated.

\par The OAMPNet2 detector outperforms the OAMP detector in all MIMO dimensions with 16QAM modulation as illustrated in \figurename{\ref{fig9}}, This is consistent with the results in \cite{C7OAMPNet2}. Exactly, the performance gain is about $4.0$ dB, $3.0$ dB, $2.5$ dB, and $2.0$ dB in $16\times16$, $32\times32$, $64\times64$, and $128\times128$ MIMO systems, respectively. As the dimension of the MIMO system increases, the performance gap between EPD and the OAMP detector becomes narrower, this is because for additive white Gaussian noise measurement channels, the EP algorithm is proven to be equivalent to the AMP algorithm in large scale system limit \cite{D3EP-AMP}. The performance gap between EPD and the OAMPNet2 detector has the similar trend, finally both detectors have almost the same performance in $128\times128$ MIMO systems. Among these detectors, MEPD has the best performance in all settings. Specifically, MEPD outperforms EPD around $2.0$ dB at SER $=10^{-2}$ and over $3.0$ dB at SER $=10^{-3}$ in $16\times16$ MIMO system. With $N_r=N_t=32$, MEPD shows similar results as  that of $N_r=N_t=16$. When the system dimension goes to $N_r=N_t=64$ and $128$, SER performance gain between MEPD and EPD is slightly reduced but still with $2$ dB lead at SER of $10^{-3}$.

When the modulation scheme is 64QAM, performance gaps between EPD and the OAMPNet2 detector are larger than that of 16QAM modulation as shown in \figurename{\ref{fig10}}. While MEPD achieves much better performance than EPD, this is because MEPD inherits the excellent performance of EPD and is optimized for practical MIMO scenarios. Overall, MEPD outperforms EPD in all settings, of which larger performance gain can be obtained in high SNR regime. Such results fully illustrate the superiority of MEPD and the effectiveness of parameter optimization using a deep learning scheme.

\par \figurename{\ref{fig11}} illustrates the SER performance of MEPD, EPD, and MMSE-SIC detector in asymmetric MIMO systems with 16QAM modulation. Both MIMO systems with $\eta=0.8$ and $\eta=0.5$ are investigated. We can find that MEPD still outperforms EPD in all settings although the performance gap is narrowing because of the effect of channel hardening \cite{D1CHEMP}.

\subsection{Correlated MIMO Channel Performance}
\begin{figure}[!t]
      \centering
      \includegraphics[width= 3.5 in]{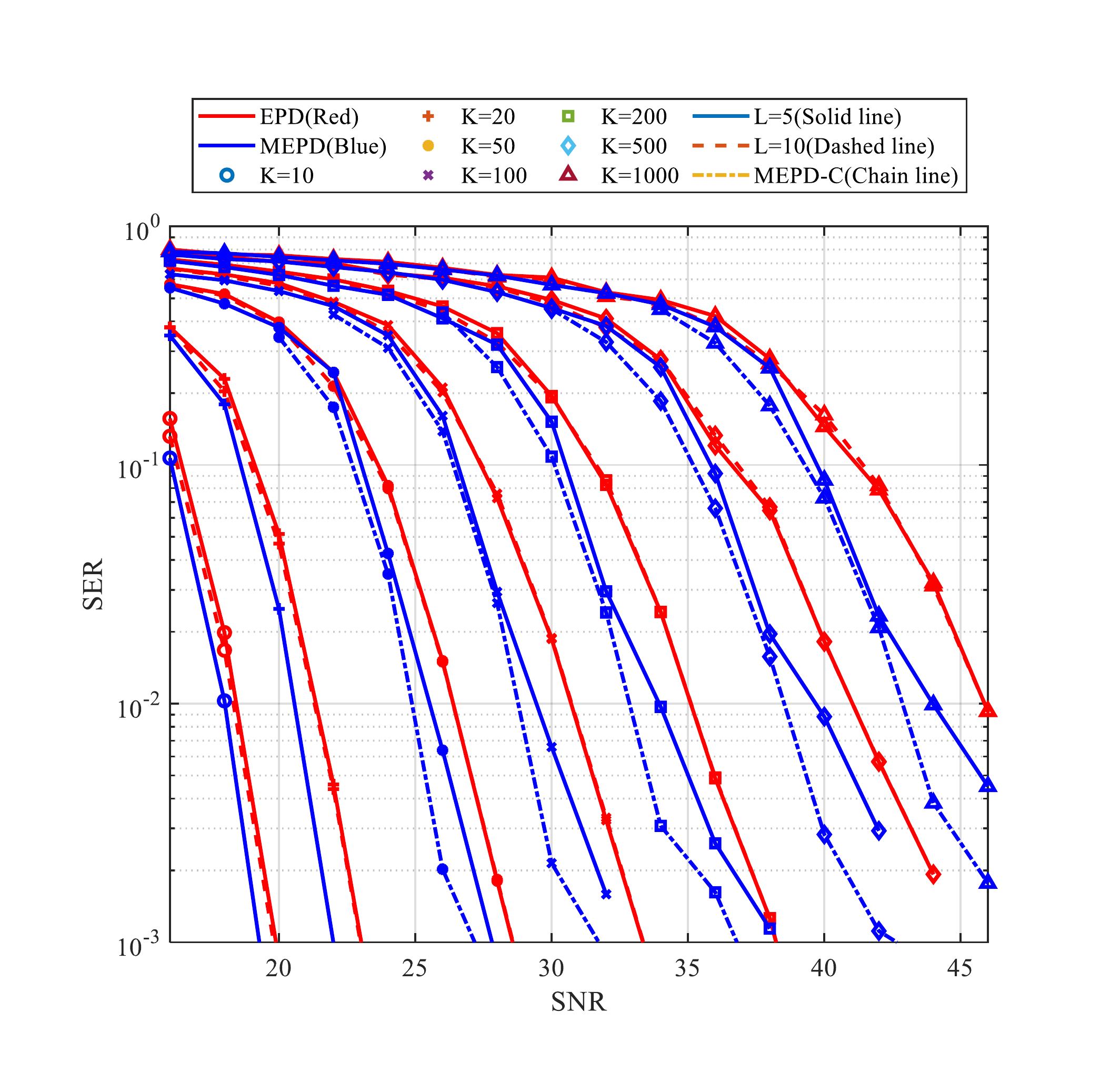}
      \caption{\small{SERs performance of the MEPD and EPD versus SNRs under $32\times32$ correlated MIMO channels with 16QAM modulation.}}
      \label{fig12}
\end{figure}

\par \figurename{\ref{fig12}} compares the SER performance of MEPD and EPD in correlated MIMO channels. Blue solid lines represent the performances of MEPD which are trained with i.i.d Rayleigh MIMO channels but tested under correlated MIMO channels, while blue chain lines named MEPD-C represent the performances of MEPD, whose parameters are trained and tested under correlated MIMO channels. The gaps between blue and red solid lines illustrate that MEPD with parameters trained under i.i.d channels still surpasses EPD in all settings, Moreover, the performance gain will be larger if MEPD is trained under correlated MIMO channels shown as blue chain lines.

\subsection{Robustness with SNR Mismatch}
\begin{figure}[!t]
      \centering
      \includegraphics[width= 3.5 in]{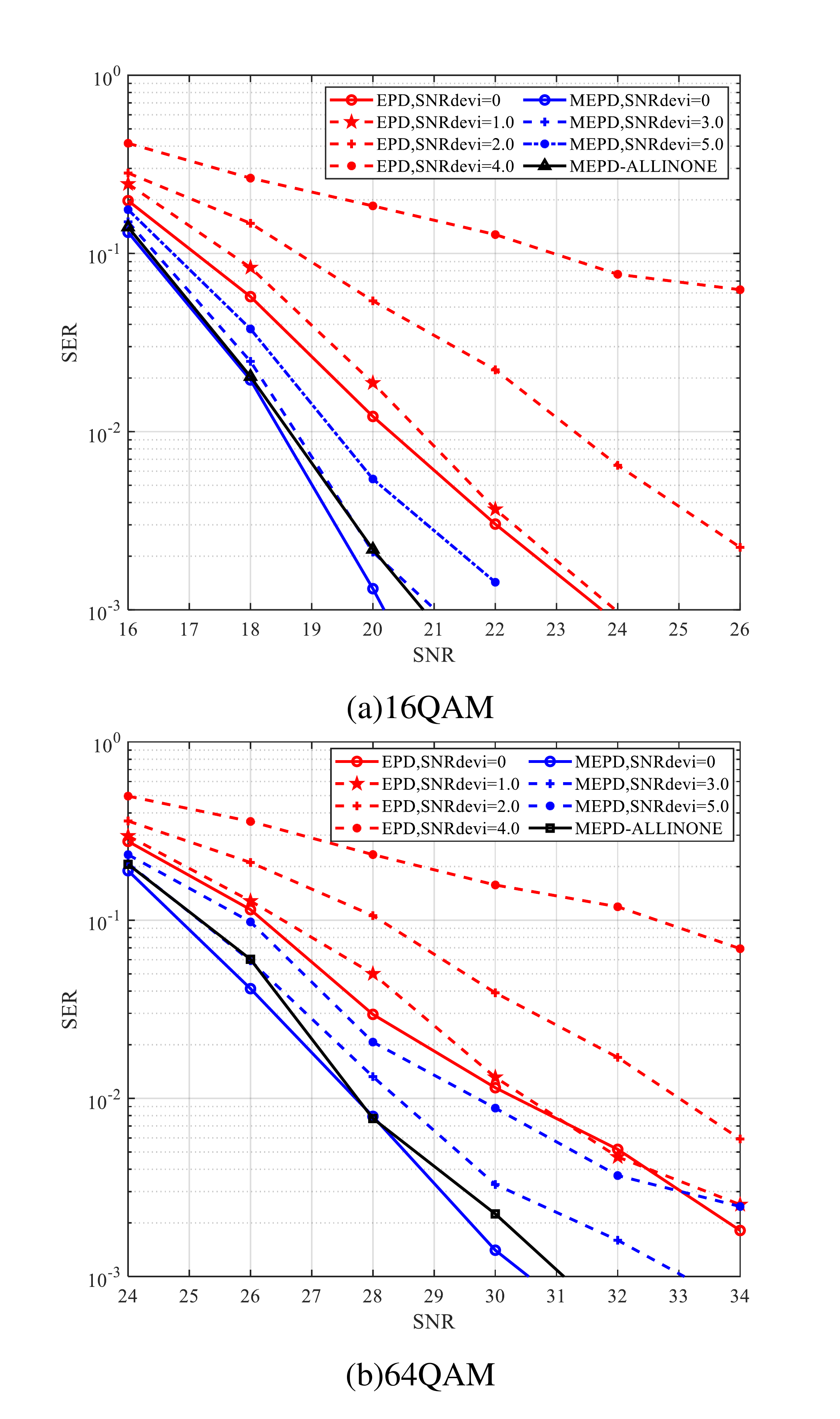}
      \caption{\small{SERs performance of the MEPD and EPD versus SNRs under 32x32 i.i.d. Rayleigh MIMO channels with (a)16QAM and (b)64QAM modulation.}}
      \label{fig13}
\end{figure}

\par In the practical implementation scheme proposed in section IV, we can find that the accuracy of the estimation of noise power is of critical importance because MEPD would load the optimized parameters according to $\hat{\sigma_n^2}$. What if the wrong parameters are loaded when the estimation error of noise power reaches $3.0$ dB and even larger? Because MEPD is trained every $2.0$ dB, if the maximum estimation error of noise power, SNRdevi, is less than $1.0$ dB, MEPD would load the right parameters, if SNRdevi exceeds $1.0$ dB, the neighboring parameters would be loaded. So when SNRdevi $=3.0$ dB and SNRdevi $=5.0$ dB, 3 and 5 consecutive parameter pairs could be loaded with equal probability, respectively.

\par In \figurename{\ref{fig13}}, we investigated the SER performance of MEPD and EPD under $32\times32$ i.i.d. Rayleigh MIMO channels with SNR mismatches. The black solid lines named MEPD-ALLINONE represent the performance of MEPD under accurate SNR trained with all tested SNRs which means that the SNR of the training data pairs is uniformly random selected in the range of $16$ dB-$26$ dB and $24$ dB-$34$ dB for 16QAM and 64QAM modulation, respectively. We can find that the performance of MEPNet-ALLINONE is nearly the same with that of MEPD under a maximum $3.0$ dB SNR mismatch. Even when SNRdevi $=5.0$ dB, MEPD still outperforms EPD with no SNR mismatch.

\section{Conclusion}
\par In this paper, we have demonstrated that EPD cannot achieve Bayes-optimal performance in the practical MIMO systems. Then a modified EP-based MIMO detection algorithm (MEPD) is proposed and optimized through a deep learning scheme. MEPD can take advantage of the self-correction ability of the EP algorithm, and achieve better performance compared with that of EPD. The modified updating scheme of moment matching of MEPD is proposed based on our studies about the self-correction ability of EP algorithm. In order to obtain the optimal parameters, we proposed the MEPNet obtained by unfolding MEPD and setting these parameters as learnable variables. The simulation results demonstrate that MEPD outperforms EPD in various scenarios. Furthermore, MEPD shows better robustness under correlated MIMO channels and SNR mismatches.

\bibliographystyle{IEEEtran}
\bibliography{Mendeley}

\end{document}